\documentclass[journal]{IEEEtran}
\usepackage{cite}

\ifCLASSINFOpdf
   \usepackage[pdftex]{graphicx}
   \graphicspath{{./figures/}}
\else
\fi
\usepackage{amsmath}
\DeclareMathAlphabet{\pazocal}{OMS}{zplm}{m}{n}
\let\Phi\varPhi

\usepackage{mathptmx}
\usepackage{bm}
\usepackage{multirow}
\usepackage{threeparttable}

\begin{document}
%
\title{Generalized Multivariable Grid-Forming Control Design for Power Converters}
%
%
%

\author{Meng Chen,~\IEEEmembership{Student Member,~IEEE,}
        Dao Zhou,~\IEEEmembership{Senior Member,~IEEE,}
        Ali Tayyebi,
        Eduardo Prieto-Araujo,~\IEEEmembership{Senior Member,~IEEE,}
        Florian D$\ddot{\rm o}$rfler,~\IEEEmembership{Senior Member,~IEEE,}
        and Frede Blaabjerg,~\IEEEmembership{Fellow,~IEEE}
}

%
%

\markboth{Journal of \LaTeX\ Class Files,~Vol.~14, No.~8, August~2015}%
{Shell \MakeLowercase{\textit{et al.}}: Bare Demo of IEEEtran.cls for IEEE Journals}
%



\maketitle

\begin{abstract}
The grid-forming converter is an important unit in the future power system with more inverter-interfaced generators. However, improving its performance is still a key challenge. This paper proposes a generalized architecture of the grid-forming converter from the view of multivariable feedback control. As a result, many of the existing popular control strategies, i.e., droop control, power synchronization control, virtual synchronous generator control, matching control, dispatchable virtual oscillator control, and their improved forms are unified into a multivariable feedback control transfer matrix working on several linear and nonlinear error signals. Meanwhile, unlike the traditional assumptions of decoupling between AC and DC control, active power and reactive power control, the proposed configuration simultaneously takes all of them into consideration, which therefore can provide better performance. As an example, a new multi-input-multi-output-based grid-forming (MIMO-GFM) control is proposed based on the generalized configuration. To cope with the multivariable feedback, an optimal and structured $\pazocal{H}_{\infty}$ synthesis is used to design the control parameters. At last, simulation and experimental results show superior performance and robustness of the proposed configuration and control.
\end{abstract}

\begin{IEEEkeywords}
Grid-forming, power converter, multiple-input-multiple-output system, feedback control, $\pazocal{H}_{\infty}$ synthesis.
\end{IEEEkeywords}

%
\IEEEpeerreviewmaketitle

\section{Introduction}

\IEEEPARstart{A}{s} more inverter-interfaced generators (IIGs) are integrated into the power system, stability issues are becoming more pronounced due to the lack of inertia and poor regulation of the frequency and voltage. To cope with these challenges, grid-forming converters can establish the frequency and voltage by themselves without relying on the power grid. The synchronization among the grid-forming converters and with the power grid is based on the power balance rather than on a phase-locked loop (PLL) like in a traditional grid-following converter. Therefore, by proper power control, grid-forming converters are able to participate in the frequency and voltage regulation and then help to enlarge the penetration of the IIGs in the power system. On the contrary, the stability of the power system may be deteriorated if the control of the grid-forming converter is not well designed \cite{Milano2018}.

So far, several grid-forming controls have been proposed, where a classification is summarized as follows. The basic one is droop control, which emulate the $p$-$f$ and $q$-$V$ droop characteristics of the synchronous generator (SG) to achieve frequency and voltage regulation \cite{Rocabert2012}. A similar idea is called power synchronization control (PSC), which directly builds the relationship between the active power and the angle like the synchronous generator as well \cite{Pan2020}. To improve the dynamic performance by limiting the fast change of the frequency, the virtual synchronous generator (VSG) control has been proposed, which not only has the droop characteristics but also can provide the favorable inertia characteristics \cite{Liu2016a} for small rate of change of frequency (RoCoF). Another interesting grid-forming control is matching control. The motivation is that the dynamics of the DC capacitor are similar to the swing equation of the synchronous generator, and the frequency can be built by the DC voltage \cite{Arghir2018}. In contrast with the aforementioned SG-emulation methods, a dispatchable virtual oscillator control (dVOC) is also proposed to treat the grid-forming converters as coupled oscillators \cite{Tayyebi2020}. More recently, a hybrid angle control (HAC) is proposed in \cite{Tayyebi2020hybrid}, which uses nonlinear controllers and combines the DC and AC dynamics to give superior performance. 

Based on the aforementioned fundamental controls, many improved methods have been reported in the literature in order to provide better performance. A robust droop control is proposed in \cite{Zhong2013}, where an additional voltage feedback is added to improve the accuracy of power sharing. Further in \cite{Peng2017}, a more additional frequency feedback is added as well. A generalized droop control is proposed by replacing the fixed gain with transfer functions in \cite{Meng2019} to improve the performance of the closed-loop system. By changing the integral (I) controller to the proportional-integral (PI) controller, the virtual inertia is included in the PSC in \cite{Vasquez2013}. In \cite{Zhang2016a}, a generalized PSC is designed, which can also integrate both the droop and inertia characteristics. In \cite{Quan2020a}, a novel PSC is proposed in order to freely adjust the damping. In \cite{Mo2017} and \cite{Shuai2020}, the high-frequency component of the frequency is used to improve the damping of the VSG. In \cite{Shuai2019} and \cite{Xiong2021}, the coupling terms of the voltage and frequency are designed to enhance the stability of the VSG, respectively. In \cite{Li2021a}, the DC voltage is also used to provide additional damping for the matching control. Other improved grid-forming controllers have been presented in \cite{Huang2018,Liu2019,Liu2020,Chen2021a,Chen2021}. Given the numerous grid-forming controllers and their variations, it is hard to compare and evaluate these methods. Therefore, it is important to study the relationships among different grid-forming methods as well as to unify and generalize their control architectures.

In \cite{Huang2020}, generalized grid-connected controls encompassing droop control have been proposed. The initial and most tries for unifying grid-forming controls are focusing on the basic droop control with low-pass filters (LPFs) and the VSG control. It has been proven that they are identical in some cases for properly selected parameters \cite{DArco2014,Liu2016a}. In \cite{Meng2019}, the droop control and the VSG are unified by analyzing different damping terms. The basic droop control and the PSC control are also shown to be identical to some extent \cite{Pan2020}. Besides, a unified modeling method is presented in \cite{Liu2020} focusing on several VSG controls from the view of frequency control. On the contrary, different voltage controls of the VSG are compared in \cite{Chen2021a}. By considering more kinds of grid-forming controllers, a comprehensive comparison is given in \cite{Tayyebi2020}, while no commonalities are highlighted. 

According to the above discussion, the existing works on unifying the grid-forming controllers have limitations in four aspects. First, most of the studies consider only the basic control architectures, while their improved variations cannot be included. Second, only one or two kinds but not all of the grid-forming controllers have been unified within in a single architecture so far. Third, only single-input-single-output (SISO) loops are considered, i.e., most of the studies assume that the DC and AC sides and the active and reactive power loops are decoupled, which is not true in reality. Fourth, manual parameter tuning of many nested single-input-single-output (SISO) loops is usually burdensome and cannot achieve the optimal control performance for a multi-input-multi-output (MIMO) system \cite{Rosso2019a}.

In order to overcome the aforementioned limitations, this paper looks at the grid-forming converter design from the perspective of modern control theory. First, the grid-forming converter is abstracted as a MIMO system. Then a multivariable feedback control architecture is proposed to provide a generalized configuration, which results in the following advantages:
\begin{enumerate}
	\item The proposed configuration unifies and generalizes different grid-forming controls - not only the basic formulations, but also many of their improved variations - in a control transfer matrix relating on several linear and nonlinear error signals.
	\item The comparisons between different grid-forming controls can be performed in a straightforward way. 
	\item Different loops, e.g., DC control, active power control, and reactive power control of the grid-forming controls can be tuned simultaneously to optimize the performance and robustness.
	\item New grid-forming controls can be derived from the proposed generalized configuration.
\end{enumerate}

To further highlight the design power of the proposed generalized configuration, a new multi-input-multi-output-based grid-forming (MIMO-GFM) control is also proposed in this paper, which improves the performance without increasing the order of the controllers compared with the existing basic control such as VSG and droop with LPF. Usually, classic control design methods such as root locus and loop shaping are used to design the parameters. However, they cannot deal with several adjustable parameters simultaneously to obtain the optimal performance especially in a MIMO system. As a solution, the $\pazocal{H}_{\infty}$ synthesis can be used, which has been proved to be effective in secondary frequency control \cite{Fathi2018}, current control of grid-following converter \cite{Kammer2019}, and voltage control of PLL or droop based grid-connected converters \cite{Huang2020}, etc. Nevertheless, no use in the MIMO-GFM converter including all the three control loops has been reported. In this paper, we present how the proposed MIMO-GFM controller can be transformed to the standard $\pazocal{H}_{\infty}$ synthesis, where the fixed-structure $\pazocal{H}_{\infty}$ synthesis is performed to optimize the parameters of the controller.

The remainder of the paper is organized as follows: Section II presents the proposed generalized configuration and unifies the existing grid-forming controllers. Section III proposes a new MIMO-GFM controller and gives the details of parameters design based on the $\pazocal{H}_{\infty}$ synthesis. Simulations and experimental results are performed in Section IV, and the conclusions are drawn in Section V.  

\section{General Configuration of Grid-Forming Converter}

\begin{figure}[!t]
	\centering
	\includegraphics[width=\columnwidth]{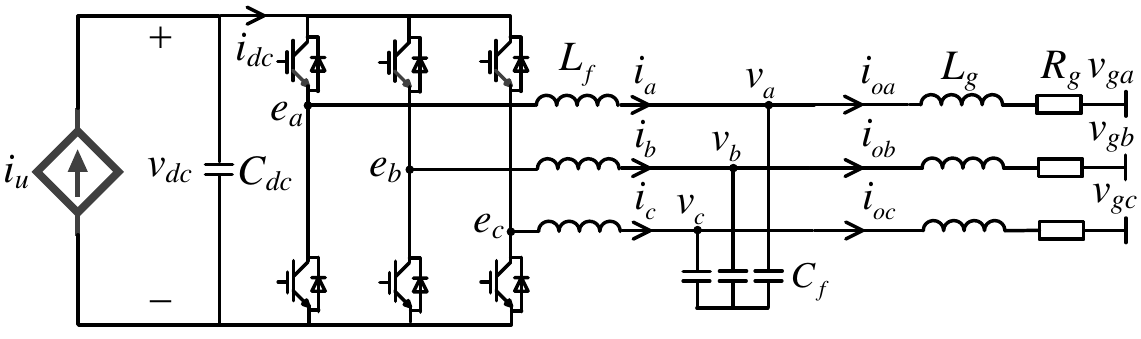}
	\caption{Topology of grid-forming converter.}
	\label{VSC}
\end{figure}

The studied topology of the grid-forming converter is shown in Fig. \ref{VSC}, where a three-phase inverter is connected to the power grid via a filter. $L_f$ and $C_f$ are the inductor and capacitor of the filter. $L_g$ and $R_g$ are the equivalent inductor and resistor to the power grid. Most of the existing works assume that the DC source is ideal and able to decouple the AC and DC sides. Although this assumption simplifies the analysis, the information in the dynamics of the DC capacitor is missing, which can be used to improve the performance. Meanwhile, some grid-forming controls such as the matching control is based on the coupling between AC and DC sides. Therefore, in this paper, the dynamics of the capacitor are included, where the DC source is equivalent to a controlled current source $i_u$ paralleled with a capacitor $C_{dc}$ \cite{Tayyebi2020}.

\subsection{MIMO Formulation of Grid-Forming Converter}

The modeling of AC side of the grid-forming converter in $d$-$q$ frame defined by the input $\omega_u$ is summarized as follows:
\begin{align}
&\dot i_d=\frac{\omega_b}{L_f}E_u-\frac{\omega_b}{L_f}v_d+\omega_b\omega_ui_q\\
&\dot i_q=-\frac{\omega_b}{L_f}v_q-\omega_b\omega_ui_d\\
&\dot v_d=\frac{\omega_b}{C_f}i_d-\frac{\omega_b}{C_f}i_{od}+\omega_b\omega_uv_q\\
&\dot v_q=\frac{\omega_b}{C_f}i_q-\frac{\omega_b}{C_f}i_{oq}-\omega_b\omega_uv_d\\
&\dot i_{od}=\frac{\omega_b}{L_g}v_d-\frac{\omega_b}{L_g}V_g\cos\delta-\frac{\omega_bR_g}{L_g}i_{od}+\omega_b\omega_ui_{oq}\\
&\dot i_{oq}=\frac{\omega_b}{L_g}v_q+\frac{\omega_b}{L_g}V_g\sin\delta-\frac{\omega_bR_g}{L_g}i_{oq}-\omega_b\omega_ui_{od}
\end{align}
where $i_{dq}$, $v_{dq}$, and $i_{odq}$ are the currents of the filter inductor, voltages of the filter capacitor, and output currents, respectively, $\omega_b$ is the base angular frequency, $V_g$ is the voltage magnitude of the power grid, $\omega_u$ and $E_u$ are the frequency and voltage provided by the grid-forming control, $\delta$ is the angle difference between the grid-forming converter and the power grid, which is defined as
\begin{align}
\dot\delta=\omega_b\omega_u-\omega_b\omega_g,
\end{align}
where $\omega_g$ is the angular frequency of the power grid. Furthermore, the DC dynamics are modeled as
\begin{align}
\dot v_{dc}=\frac{\omega_b}{C_{dc}}i_u-\frac{\omega_bE_ui_d}{C_{dc}v_{dc}},
\end{align}
where $v_{dc}$ is the DC voltage. It is noted that the DC dynamics contain the information of the AC side by $E_ui_d$, which provides the possibility to improve the AC performance by the DC signals. We stress that, the usual assumption of ideal DC source is not beneficial to optimize the grid-forming converter.

For the grid-forming converter, five outputs are usually considered, i.e., the active and reactive power $p$ and $q$, magnitude of the terminal voltage $V$, the frequency assigned by the grid-forming control $\omega_u$, as well as $v_{dc}$, where $p$, $q$, $V$ are expressed by the state variables as follows:
\begin{align}
&p=v_di_{od}+v_qi_{oq}\\
&q=-v_di_{oq}+v_qi_{od}\\
&V=\sqrt{v^2_d+v^2_q}
\end{align}

\begin{figure}[!t]
	\centering
	\includegraphics[width=\columnwidth]{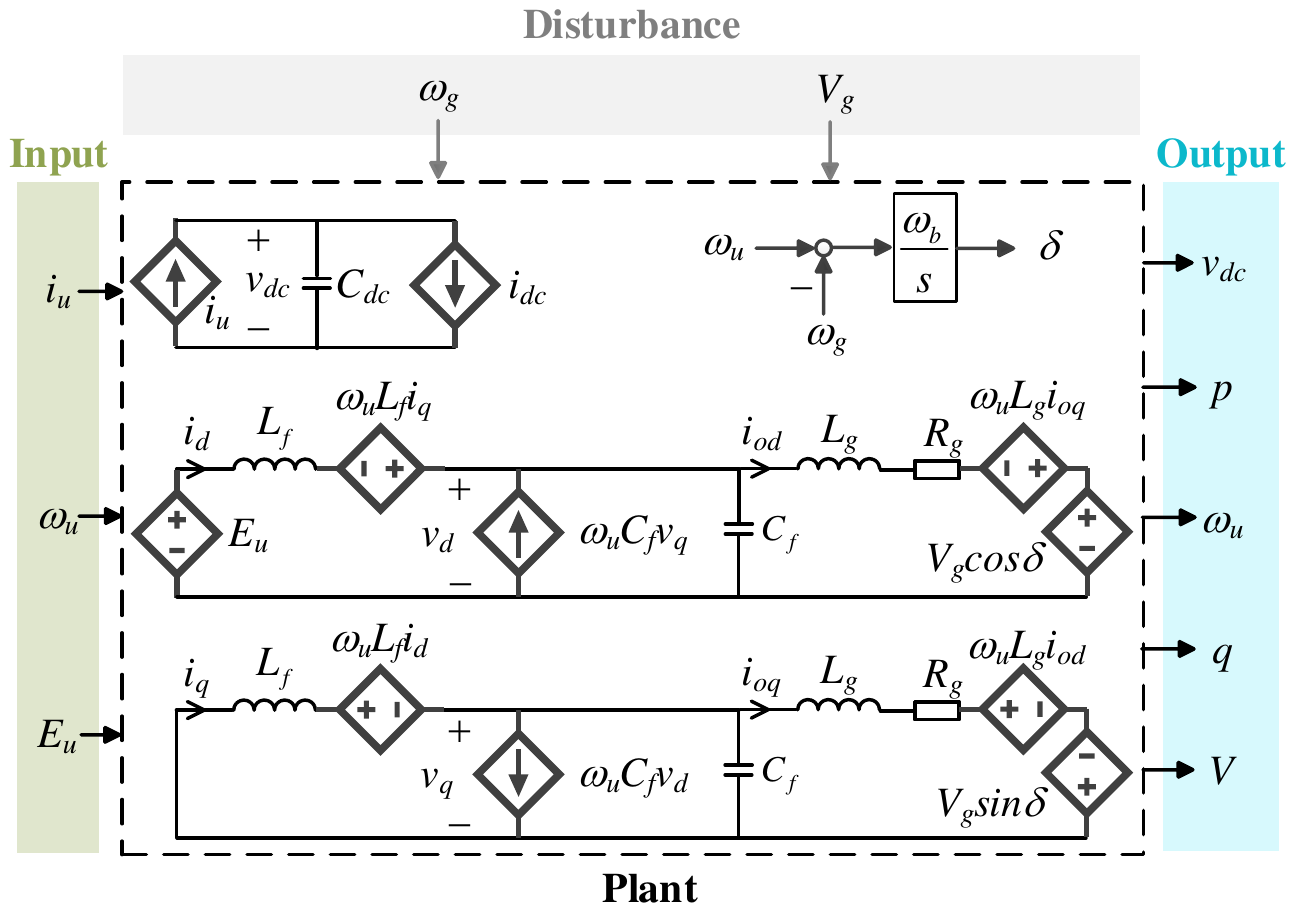}
	\caption{MIMO open-loop equivalent circuit of grid-forming converter in $d$-$q$ frame.}
	\label{equivalent}
\end{figure}

Hence, the open-loop equivalent circuit of the grid-forming converter can be represented by a MIMO system, as shown in Fig. \ref{equivalent}. There are three control inputs, which should be provided by the grid-forming control, and five outputs, which should be regulated. Furthermore, the grid voltage and frequency can be seen as disturbances.

For notational convenience, we define the vectors:
\begin{align}
&\bm x=\left[
	\begin{matrix}
	i_d&i_q&v_d&v_q&i_{od}&i_{oq}&\delta&v_{dc}
	\end{matrix}
\right]^T\\
&\bm u=\left[
	\begin{matrix}
	i_u&\omega_u&E_u
	\end{matrix}
\right]^T\\
&\bm y=\left[
	\begin{matrix}
	v_{dc}&p&\omega_u&q&V
	\end{matrix}
\right]^T\\
&\bm d=\left[
	\begin{matrix}
	\omega_g&V_g
	\end{matrix}
\right]^T,
\end{align}
where $\bm x$ is the state vector, $\bm u$ is the control vector, $\bm y$ is the output vector, and $\bm d$ is the disturbance vector. Thereafter, the equivalent circuit of Fig. \ref{equivalent} can be abstracted as the open-loop state-space model as shown in Fig. \ref{open_control}, from which the target of the grid-forming control is defined as the following:

\begin{figure}[!t]
	\centering
	\includegraphics[width=0.45\columnwidth]{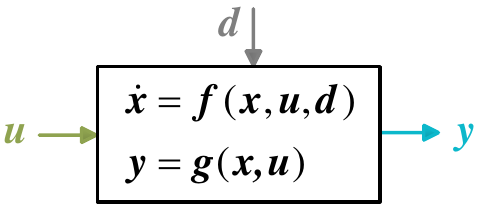}
	\caption{MIMO open-loop state-space model of grid-forming converter.}
	\label{open_control}
\end{figure}

\textit{A grid-forming control is to design a control input \textbf{u} for the system (\textbf{f}, \textbf{g}) to guarantee that the output \textbf{y} satisfies the desired specifications (details will be given later in Section III) in the presence of disturbance \textbf{d}.}

\begin{figure}[!t]
	\centering
	\includegraphics[width=0.9\columnwidth]{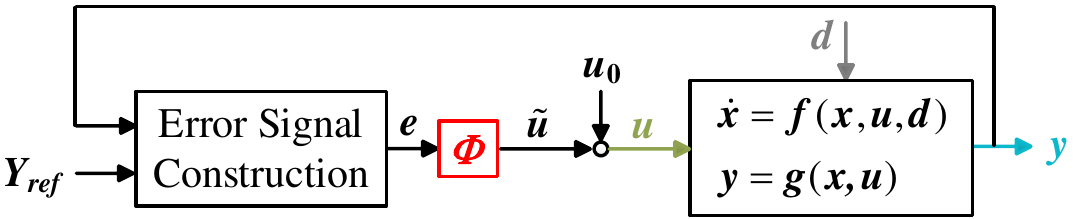}
	\caption{General MIMO close-loop feedback control configuration of grid-forming converter.}
\label{close_control}
\end{figure}

To cope with this problem, a multivariable feedback control can be used to close the loop. Therefore, a generalized configuration of the grid-forming converter control architecture is proposed, as shown in Fig. \ref{close_control},
\begin{align}
&\bm{u_0}=\left[
	\begin{matrix}
	i_0&\omega_0&E_0
	\end{matrix}
\right]^T\\
&\bm{Y_{ref}}=\left[
	\begin{matrix}
	V_{dcref}&P_{ref}&\omega_g&Q_{ref}&V_{ref}
	\end{matrix}
\right]^T\\
&\bm{\Phi}=\left(\phi_{ij}\right)_{3\times 5},
\end{align}
where $\bm{u_0}$ is the vector of set-points for $\bm u$, $\bm{Y_{ref}}$ is the vector of references for $\bm y$. We notice that the nominal reference of the frequency is chosen as $\omega_g$ due to the fact that the frequency of the grid-forming converter should be synchronized to the grid frequency in the nominal steady-state. Besides, $\bm{\Phi}=\bm{\Phi}(s)$ is the $3\times5$ control transfer matrix, and $\bm e$ is the vector of error signals. The achievement of the grid-forming controller is therefore attributed to the choices of $\bm{\Phi}$ and $\bm e$. A simple and natural choice of $\bm e$ is using the linear error signals, i.e., $\bm e=\bm{Y_{ref}}-\bm y$. Nevertheless, some nonlinear error signals can also prove useful, which will be discussed later.

To highlight the advantages, the following section will discuss how the existing methods are unified by the proposed generalized configuration of the grid-forming converter.
\subsection{Discussion on Existing Grid-Forming Converters}
\subsubsection{Droop Control}
Droop control emulates the droop characteristics of the regulation of SG, which can be expressed as \cite{Rocabert2012,Pan2020}
\begin{align}
\label{droop_frequency}
&\omega_u-\omega_{0}=-D_p\left(p-P_{ref}\right)\\
&E_u-E_{0}=-D_q\left(q-Q_{ref}\right),
\end{align}
where $D_p$ and $D_q$ are the droop coefficients. As mentioned before, the general strategy the droop control assumes an ideal DC source and therefore neglect the DC control. This paper removes this assumption. In practice, a proportional integral (PI) controller is usually used to control the voltage of the DC capacitor, which is expressed as
\begin{align}
\label{droop_dc}
i_u=i_{0}+k_{pdc}\left(V_{dcref}-v_{dc}\right)+k_{idc}\int\left(V_{dcref}-v_{dc}\right)dt,
\end{align}
where $k_{pdc}$ and $k_{idc}$ are the proportional and integral gains, respectively.

\begin{figure}[!t]
	\centering
	\includegraphics[width=0.7\columnwidth]{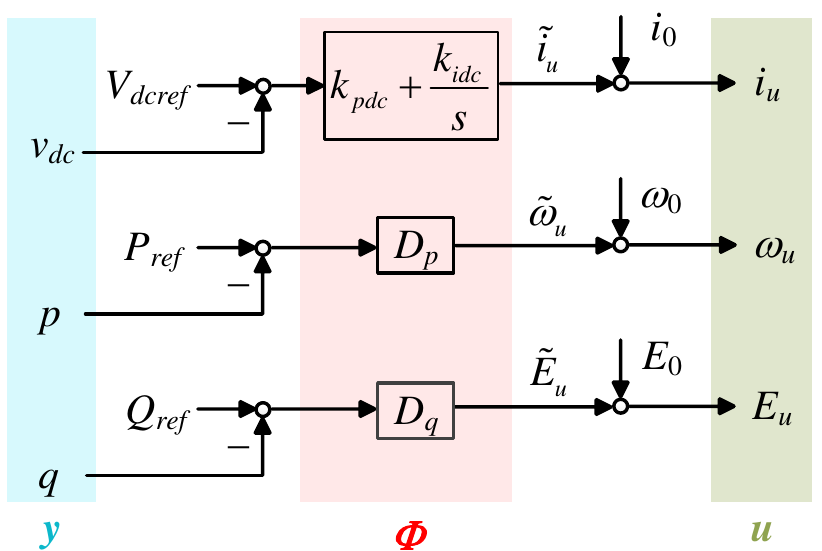}
	\caption{Block diagram of droop control.}
	\label{droop}
\end{figure}

According to (\ref{droop_frequency})-(\ref{droop_dc}), the equivalent control block of the droop control using the generalized configuration is shown in Fig. \ref{droop}, where the control transfer matrix is
\begin{align}
\label{droop_phi}
\bm\Phi=\left[
	\begin{matrix}
	k_{pdc}+k_{idc}/s&0&0&0&0\\
	0&D_p&0&0&0\\
	0&0&0&D_q&0
	\end{matrix}
\right].
\end{align}

The control transfer matrix of droop control (\ref{droop_phi}) uses a SISO PI controller and two SISO proportional (P) controllers to derive the grid-forming control input, and there is no coupling between different columns and rows. Therefore, effective ($v_{dc}$, $p$, $\omega_u$, $q$, $V$) coupling that exists in reality is typically not considered.

\subsubsection{PSC Control}
The PSC control directly emulates the relationship between the rotor angle and the active power of the SG, which can be expressed as \cite{Pan2020}
\begin{align}
D_p\int\left(P_{ref}-p\right)dt+\omega_{0}t=\int\omega_udt
\end{align}
which can be rewritten in the form of the general configuration, by solving for $\omega_u$, as
\begin{align}
\label{psc_frequency}
D_p(P_{ref}-p)+\omega_{0}=\omega_u
\end{align}

By comparing (\ref{psc_frequency}) with (\ref{droop_frequency}), we conclude that PSC control is equivalent to droop control with respect to $\omega_u$. Moreover, the DC and AC voltage controls of PSC control can also be chosen to be identical as in droop control. Therefore, the control transfer matrix of the PSC control has the same structure as droop control in (\ref{droop_phi}).

\subsubsection{VSG Control}

\begin{figure}[!t]
	\centering
	\includegraphics[width=0.9\columnwidth]{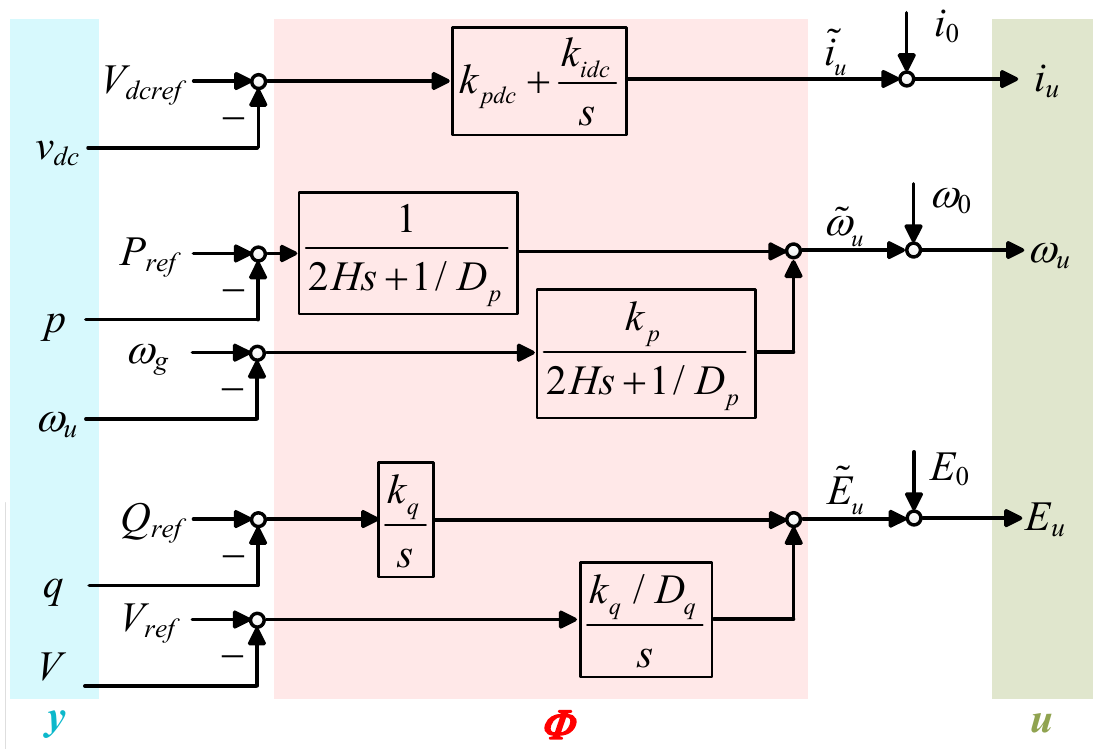}
	\caption{Block diagram of VSG control.}
	\label{VSG}
\end{figure}

The VSG control emulate the SG in a more detailed fashion than droop control and PSC control. One of basic architectures can be expressed as \cite{Liu2020,Chen2021a,Rodr2019}
\begin{align}
&2H\Delta\dot\omega_u=P_{ref}-p-\frac{1}{D_p}\left(\omega_u-\omega_{0}\right)-k_p(\omega_u-\omega_g)\\
&k_q\int\left[Q_{ref}-q+\frac{1}{D_q}\left(V_{ref}-V\right)\right]dt+E_{0}=E_u,
\end{align}
where the inertia constant $H$ emulates the inertia characteristics and the damping coefficient $k_p$ emulates the damping characteristics. Furthermore, the DC control is usually based on a PI control, as in (\ref{droop_dc}). Therefore, the equivalent control block of the VSG control using the generalized configuration is shown in Fig. \ref{VSG}, where the control transfer matrix is
\begin{align}
\label{VSG_phi}
\bm\Phi=\left[
	\begin{matrix}
	k_{pdc}+k_{idc}/s&0&0&0&0\\
	0&\frac{1}{2Hs+1/D_p}&\frac{k_p}{2Hs+1/D_p}&0&0\\
	0&0&0&k_q/s&\frac{k_q/D_q}{s}
	\end{matrix}
\right].
\end{align}

The control transfer matrix (\ref{VSG_phi}) of the VSG uses PI controller, low-pass filters (LPFs), and I controllers to derive the control inputs, respectively, and there are again no coupling terms among different rows, i.e., the couplings among DC voltage, AC voltage, and frequency controls are not considered.

\subsubsection{Matching Control}

\begin{figure}[!t]
	\centering
	\includegraphics[width=0.7\columnwidth]{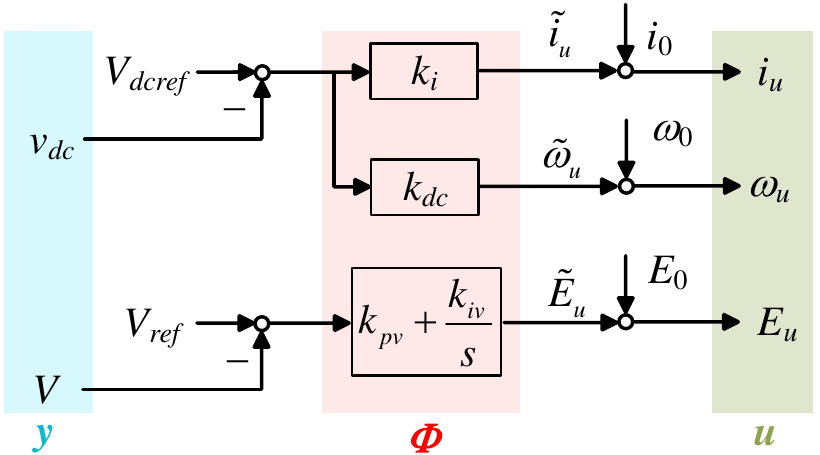}
	\caption{Block diagram of matching control.}
	\label{matching}
\end{figure}

The matching control is motivated by the fact that the dynamics of the DC capacitor are similar to the rotor in the SG, and therefore $\omega_u$ can be generated by $v_{dc}$. The mathematical model of the matching control can be expressed as \cite{Arghir2018}
\begin{align}
&i_u=i_{0}+k_i(V_{dcref}-v_{dc})\\
&\omega_u=\omega_{0}+k_{dc}\left(V_{dcref}-v_{dc}\right)\\
&k_{pv}\left(V_{ref}-V\right)+k_{iv}\int\left(V_{ref}-V\right)dt+E_{0}=E_u
\end{align}
based on which the equivalent control block using the generalized configuration is shown in Fig. \ref{matching}. Thereafter, the control transfer matrix of the matching control can be derived as
\begin{align}
\bm\Phi=\left[
	\begin{matrix}
	k_i&0&0&0&0\\
	k_{dc}&0&0&0&0\\
	0&0&0&0&k_{pv}+k_{iv}/s
	\end{matrix}
\right].
\end{align}

The matrix of the matching control uses two P controllers and one PI controller to derive the control inputs. Note that the first two rows are coupled via the DC signal.

The above analysis shows how a selection of the basic grid-forming controllers can be transformed to the proposed generalized configuration. By analogous reasoning also their various improved formulations can be represented as particular control transfer matrices. Table \ref{table_matrix} summarizes the control transfer matrices employed in different grid-forming controllers.

\begin{table*}[!t]
\renewcommand{\arraystretch}{1.3}
\caption{Summary of Control Transfer Matrices Corresponding to Different Grid-forming Controllers}
\label{table_matrix}
\centering
\begin{threeparttable}
\begin{tabular}{c|ccccc|ccccc|ccccc}
\hline\hline
Feedback Signals $\bm y$&$v_{dc}$&$p$&$\omega_u$&$q$&$V$&$v_{dc}$&$p$&$\omega_u$&$q$&$V$&$v_{dc}$&$p$&$\omega_u$&$q$&$V$\\\hline\hline
Transfer Matrix $\phi_{ij}$&$\phi_{11}$ & $\phi_{12}$& $\phi_{13}$& $\phi_{14}$& $\phi_{15}$& $\phi_{21}$& $\phi_{22}$& $\phi_{23}$& $\phi_{24}$& $\phi_{25}$& $\phi_{31}$& $\phi_{32}$& $\phi_{33}$& $\phi_{34}$& $\phi_{35}$\\
\hline
droop-1 \cite{Rocabert2012,Pan2020} & PI & 0 & 0 & 0 & 0 & 0 & P & 0 & 0&0&0&0&0&P&0\\
droop-2 \cite{Zhong2013} & PI & 0 & 0 & 0 & 0 & 0 & 0 & 0 & P&0&0&I&0&0&I\\
droop-3 \cite{Pan2020,Fu2021} & PI & 0 & 0 & 0 & 0 & 0 & IF & 0 & 0&0&0&0&0&0&0\\
droop-4 \cite{Vasquez2013} & PI & 0 & 0 & 0 & 0 & 0 & PD & 0 & 0&0&0&0&0&P&0\\
droop-5 \cite{Liu2016a} & PI & 0 & 0 & 0 & 0 & 0 & P\{IF$\times$D\} & 0 & 0&0&0&0&0&P&0\\
PSC-1 \cite{Pan2020} & PI & 0 & 0 & 0 & 0 & 0 & P & 0 & 0&0&0&0&0&P&0\\
PSC-2 \cite{Zhang2016a} & PI & 0 & 0 & 0 & 0 & 0 & IF$\times$PD & 0 & 0&0&0&0&0&I&0\\
PSC-3 \cite{Quan2020a} & PI & 0 & 0 & 0 & 0 & 0 & IF$\times$PD & 0 & 0&0&0&0&0&I&0\\
VSG-1 \cite{Chen2021a,Chen2020b} & PI & 0 & 0 & 0 & 0 & 0 & IF & 0 & 0&0&0&0&0&0&IF\\
VSG-2 \cite{Liu2020,Chen2021a,Rodr2019} & PI & 0 & 0 & 0 & 0 & 0 & IF & 0 & 0&0&0&0&0&PI&PI\\
VSG-3 \cite{Xiong2021} & PI & 0 & 0 & 0 & 0 & 0 & IF & 0 & 0&0&0&0&P&P&0\\
VSG-4 \cite{Xiong2021a} & PI & 0 & 0 & 0 & 0 & 0 & IF & IF & 0&0&0&0&0&P&0\\
VSG-5 \cite{Liu2016a,Liu2020,Cheng2020} & PI & 0 & 0 & 0 & 0 & 0 & IF & IF & 0&0&0&0&0&PI&PI\\
VSG-6 \cite{Shuai2019} & PI & 0 & 0 & 0 & 0 & 0 & IF & 0 & 0&IF&0&0&0&I&I\\
VSG-7 \cite{Shuai2020} & PI & 0 & 0 & 0 & 0 & 0 & O$\times$PD & 0 & 0&0&0&0&0&I&I\\
VSG-8 \cite{Liu2016a} & PI & 0 & 0 & 0 & 0 & 0 & IF$\times$PD & 0 & 0&0&0&0&0&PI&PI\\
VSG-9 \cite{Huang2018} & PI & 0 & 0 & 0 & 0 & IF$\times$PD & IF$\times$PD & 0 & 0&0&0&0&0&P&0\\
VSG-10 \cite{Liu2020} & PI & 0 & 0 & 0 & 0 & 0 & IF$_1$\{IF$_1\times$IF$_2\times$D\} & 0 & 0&0&0&0&0&PI&PI\\
VSG-11 \cite{Chen2021} & PI & 0 & 0 & 0 & 0 & 0 & O$\times$PD\{O$\times$IF$\times$PD$\times$D\} & 0 & 0&0&0&0&0&I&I\\
VSG-12 \cite{Liu2020,Liu2019} & PI & 0 & 0 & 0 & 0 & 0 & O$\times$PD$_1$\{O$\times$IF$\times$PD$_2\times$D\} & 0 & 0&0&0&0&0&PI&PI\\
matching-1 \cite{Arghir2018} & P & 0 & 0 & 0 & 0 & P & 0 & 0 & 0&0&0&0&0&0&PI\\
matching-2 \cite{Li2021a} & 0 & 0 & 0 & 0 & 0 & P & 0 & 0 & 0&0&P&0&0&0&0\\
\hline\hline
Generated Inputs $\bm u$&\multicolumn{5}{c|}{$i_u$}&\multicolumn{5}{c|}{$\omega_u$}&\multicolumn{5}{c}{$E_u$}\\\hline\hline
\end{tabular}
\begin{tablenotes}
	\footnotesize
\item P: Proportional controller $k$, I: Integral controller $\frac{1}{Ts}$, D: Derivative controller $Ts$, PI: Proportional integral controller $k(1+\frac{1}{Ts})$, PD: Proportional derivative controller $k(1+Ts)$, IF: Inertia factor $\frac{k}{Ts+1}$, O: Oscillatory factor $\frac{k}{T^2s^2+2T\xi s+1}$.
\item \{\}: the term is only applied to the feedback channel.
\end{tablenotes}
\end{threeparttable}
\end{table*}

\subsection{Discussion on Proposed Generalized Configuration}
According to the aforementioned analysis, the following advantages of the proposed generalized configuration of the grid-forming converter can be concluded:
\begin{enumerate}
	\item Not only the basic formulations of the grid-forming controls, but also many of their improved variations can be presented as control transfer matrices in a unified setting.
	\item The comparisons between different grid-forming controls is straightforward. From Table \ref{table_matrix}, it can be deduced how the performance may improve by changing the elements of the control transfer matrix. A typical strategy is using higher-order controllers, especially in the frequency control of deriving $\omega_u$. Meanwhile, the relationships between different controls are obvious. For example, although PSC-2, PSC-3, and VSG-8 have distinctions from the original control blocks, and they are derived from different motivations, their frequency controls are actually identical. As another example, many works have proved the equivalence between droop-5 and VSG-2 in the frequency control \cite{Liu2016a,DArco2014}. However, this is not entirely correct from Table \ref{table_matrix}. When the disturbance is from the output side, e.g., $p$, the droop-5 is actually identical with VSG-2. However, if the disturbance is from the input side, e.g., $P_{ref}$, they are not identical. A similar analysis can be applied to study also other methods.
	\item Different loops, i.e., DC control, active power control, and reactive power control of the grid-forming controls can be taken care of simultaneously to optimize the performance. Most of the existing grid-forming controls aim to decouple those control loops to simplify the design and analysis. However, note from Table \ref{table_matrix} that some methods have successfully used some coupling terms to improve the performance.
	\item New grid-forming controls can be inspired: In the proposed generalized configuration, the design of the grid-forming control is attributed to the control transfer matrix $\Phi$. In the future, two directions can be pursued to propose new grid-forming controls. On the one hand, most of the used elements of the existing controls are linear. Nonlinear controllers such as sine function may be used especially to improve the global stability like in \cite{Tayyebi2020hybrid}. On the other hand, the control transfer matrix $\Phi$ of the existing controls are quite sparse and different coupling terms, e.g., the DC signals motivated from the matching control \cite{samanta2021stability}, can be added. In the following, we will present an example design.
\end{enumerate}

It is worth to mention that the proposed generalized configuration can be further generalized in the following aspects:
\begin{enumerate}
	\item This paper does not consider the virtual impedance and inner voltage and current loops, as they are not the essence of the grid-forming functions and not appear in some controls \cite{Liu2016a,Arghir2018,Shuai2020}. Nevertheless, they can be included by just adding their equations in ($\bm f$, $\bm g$) in Fig. \ref{close_control}. Moreover, the multivariable control design can be used to replace these SISO loops.
	\item As mentioned before, the linear error $\bm e=\bm{Y_{ref}}-\bm y$ is a simple and natural choice. Nevertheless, other kinds of error signals can also be used, where, for example, $(v^2_{dcref}-v^2_{dc})/2$ is used in PSC in \cite{Zhang2010}. Furthermore, $P_{ref}/V^2_{ref}-p/V^2$ and $Q_{ref}/V^2_{ref}-q/V^2$ can be used as well, which is the case of dVOC \cite{Tayyebi2020}. Thus, by enlarging the available error signals, more kinds of grid-forming control can be unified into the proposed configuration.
	\item The control transfer matrices of grid-forming controls summarized in Table \ref{table_matrix} are all with fixed parameters. They can also be chosen as time-varying or functions of the angle of the (possibly estimated) grid impedance to encompass various kinds of adaptive control \cite{Vasquez2009}.
\end{enumerate}

\section{$\pazocal{H}_{\infty}$ Control Design of Grid-Forming Converter}
The proposed generalized configuration considers the grid-forming converter as a MIMO system with a multivariable control transfer matrix ruling out traditional design and tuning methods for SISO systems. In this section, a new structured and multivariable control transfer matrix for the MIMO-GFM control is first proposed. Then we present how the design of the control transfer matrix can be transformed into a standard fixed-structure $\pazocal{H}_{\infty}$ optimization problem, where all control parameters can be tuned simultaneously.

\subsection{Proposed Control Transfer Matrix}

As mentioned before, the existing control transfer matrices are based on SISO loops and thus sparse. Therefore, our proposed control transfer matrix tries to improve the performance by adding some coupling terms. The following possible principles are considered in this paper:
\begin{enumerate}
	\item $\phi_{11}$ is chosen as a PI controller to regulate $v_{dc}$ with a zero steady-state error as
\begin{align}
\phi_{11}=k_{pdc}+\frac{k_{idc}}{s}.
\end{align}
Nevertheless, the choice $k_{idc}=0$ is possible, which is required for the matching controllers. 
	\item $\phi_{22}$ should provide some inertia for the frequency control and maintain the prescribed steady-state droop characteristics. We thus choose it as
\begin{align}
\phi_{22}=\frac{D_pk_{22}}{s+k_{22}}.
\end{align}
	\item A PLL (i.e., $\omega_g$) is not expected to be used. Therefore the third column is chosen to be zero
\begin{align}
\phi_{13}=\phi_{13}=\phi_{33}=0.
\end{align}
This is just one option but others are possible if a PLL is available to improve the control performance.
	\item The I control is used to keep the steady-state $q$-$V$ droop characteristics as follows:
\begin{align}
\phi_{34}=D_q\phi_{35}=\frac{k_{34}}{s}.
\end{align}
	\item The coupling terms of $q$ and $V$ on the frequency control should not change the steady-state $p$-$f$ droop characteristics. Therefore, the following elements are chosen
\begin{align}
\phi_{24}=D_q\phi_{25}=k_{24}
\end{align}
	\item All the other elements are chosen, for simplicity, as P controllers to be designed.
\end{enumerate}

Overall, the proposed control transfer matrix for the MIMO-GFM takes the following form:
\begin{align}
\bm\Phi=\left[
	\begin{matrix}
	k_{pdc}+k_{idc}/s&k_{12}&0&k_{14}&k_{15}\\
	k_{21}&D_pk_{22}/(s+k_{22})&0&k_{24}&k_{24}/D_q\\
	k_{31}&k_{32}&0&k_{34}/s&\frac{k_{34}/D_q}{s}
	\end{matrix}
\right].
\end{align}
Observed that, unlike previously proposed variations and improved methods, the order of the proposed control transfer matrix $\bm\Phi$ is not increased compared to the basic droop control with LPFs and VSG, since the added coupling terms are all P controllers. The above shows one reasonable choice for $\Phi$. Other choices are possible as well.

\subsection{Parameters Design Based on $\pazocal{H}_{\infty}$ Optimization}
\subsubsection{Formulation of $\pazocal{H}_{\infty}$ synthesis}
\begin{figure}[!t]
	\centering
	\includegraphics[width=\columnwidth]{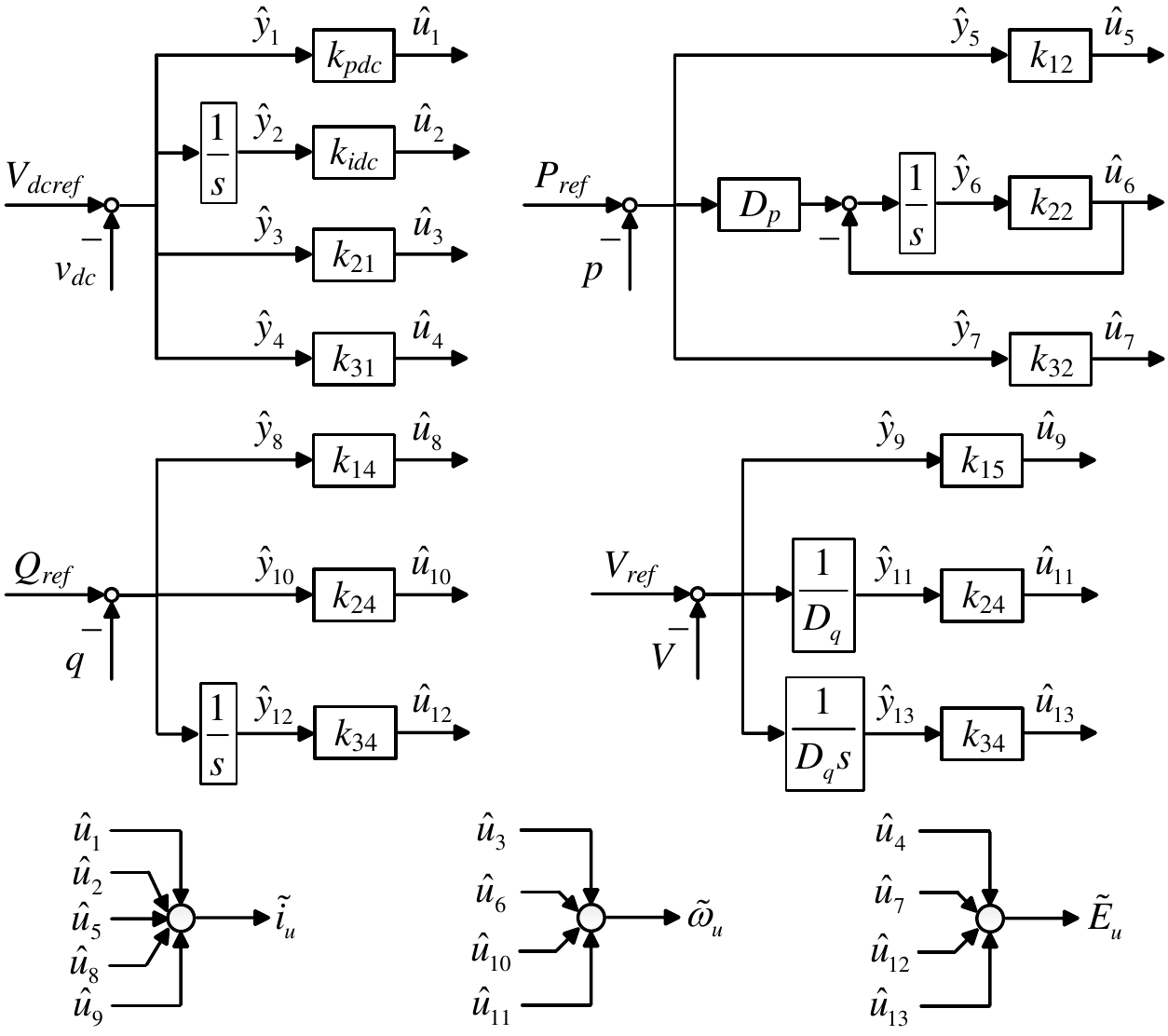}
	\caption{Block diagram of used control transfer matrix.}
	\label{h_infinity_formulation}
\end{figure}

In order to transform the control parameter design into a standard $\pazocal{H}_{\infty}$ optimization problem, the parameters to be selected are separated by defining two intermediate vectors $\bm{\hat u}$ and $\bm{\hat y}$ as shown in Fig. \ref{h_infinity_formulation}. Thus, there is the following relationship
\begin{align}
\hat{\bm u}=&diag(k_{pdc},k_{idc},k_{21},k_{31},k_{12},k_{22},k_{32},k_{14},k_{15},k_{24}\bm{I_2},k_{34}\bm{I_2})\hat{\bm y}\notag\\
=&\bm K\hat{\bm y}
\end{align}

\begin{figure}[!t]
	\centering
	\includegraphics[width=0.4\columnwidth]{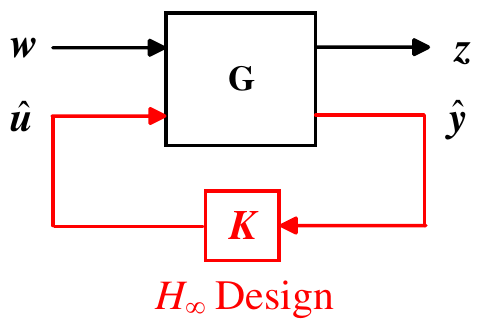}
	\caption{Block diagram of grid-forming converter in linear fractional transformation.}
	\label{LFT}
\end{figure}

It is observed that all the parameters to be designed are included in the static gain $\bm K$. Thereafter, the standard structure in the so-called linear fractional transformation for $\pazocal{H}_{\infty}$ synthesis can be derived as in Fig. \ref{LFT}, where the system in Fig. \ref{close_control} is collapsed into \textbf{G} (except for $\bm K$). Meanwhile, $\bm w$ and $\bm z$ are defined disturbance inputs and performance outputs for the $\pazocal{H}_{\infty}$ synthesis. In this paper, they are chosen as
\begin{align}
&\bm w=\left[
	\begin{matrix}
	P_{ref}&\omega_g
	\end{matrix}
\right]^T\\
&\bm z= \left[
	\begin{matrix}
	P_{ref}-p&p&\omega_u&q+V/D_q
	\end{matrix}
\right]^T
\end{align}

\subsubsection{Choices of Weighting Functions}
The $\pazocal{H}_{\infty}$ optimization uses some weighting functions to limit the disturbance responses from $\bm w$ to $\bm z$. Let $T_{ij}(s)$ be the transfer function from $w_j$ to $z_{i}$ and $W_{ij}(s)$ be the corresponding weighting function. In this paper, the following weightings $W_{ij}(s)$ are selected.

To have a small active power tracking error, low-frequency gain of $T_{11}(s)$ should be small. Therefore, $W_{11}(s)$ is chosen as:
\begin{align}
W_{11}(s)=\frac{s+s_{11\_1}}{s+s_{11\_2}}.
\end{align}

To limit the high-frequency disturbance to the active power, high-frequency gain of $T_{21}(s)$ and $T_{22}(s)$ should rapidly decay. Therefore, $W_{21}(s)$ and $W_{22}(s)$ are chosen as:
\begin{align}
&W_{21}(s)=\left({\frac{T_{21\_1}s+1}{T_{21\_2}s+1}}\right)^2\\
&W_{22}(s)=\frac{1}{k_{w22}}\times\frac{T_{22\_1}s+1}{T_{22\_2}s+1}
\end{align}

Meanwhile, the frequency of the grid-forming converter $\omega_u$ should not vary quickly as a response to $\bm w$. Therefore, the following weights $W_{31}(s)$ and $W_{32}(s)$ are chosen to limit the high-frequency gain of $T_{31}(s)$ and $T_{32}(s)$:
\begin{align}
&W_{31}(s)=\frac{1}{k_{w31}}\times\frac{s}{T_{31\_2}s+1}\\
&W_{32}(s)=\frac{T_{32\_1}s+1}{T_{32\_2}s+1}
\end{align}

Last, as mentioned before, the control of $\omega_u$ should not influence the steady-state $q$-$V$ droop regulation. Therefore, $W_{41}(s)$ is used to limit the low-frequency gain of $T_{41}(s)$ as
\begin{align}
W_{41}=\frac{s+s_{41\_1}}{s+s_{41\_2}}.
\end{align}

The used numbers in the weighting functions are summarized in Table \ref{weighting_parameter}, which are chosen depending on the converter parameters like shown in Table \ref{parameter}.

\begin{table}[!t]
	\renewcommand{\arraystretch}{1.3}
	\caption{Numbers Used in Weighting Functions}
	\centering
	\label{weighting_parameter}
	\resizebox{0.8\columnwidth}{!}{
		\begin{tabular}{cc||cc}
			\hline\hline \\[-4mm]
			Symbol & Value & Symbol & Value \\ \hline
			$s_{11\_1}$  & 4 &$s_{11\_2}$& 0.0004 \\
			$T_{21\_1}$&$1.447\times 10^{-3}$&$T_{21\_2}$&$1.447\times 10^{-5}$\\
			$k_{w22}$&100&$T_{22\_1}$&$1.447\times 10^{-3}$\\
			$T_{22\_2}$&$1.447\times 10^{-5}$&$k_{w31}$&0.015\\
			$T_{31\_2}$&$1.447\times 10^{-5}$&$T_{32\_1}$&$1.447\times 10^{-3}$\\
			$T_{32\_2}$&$1.447\times 10^{-5}$&$s_{41\_1}$  & 60\\
			$s_{41\_2}$  & 0.006\\[1.4ex]
			\hline\hline
		\end{tabular}
	}
\end{table}

\begin{figure}[!t]
	\centering
	\includegraphics[width=0.9\columnwidth]{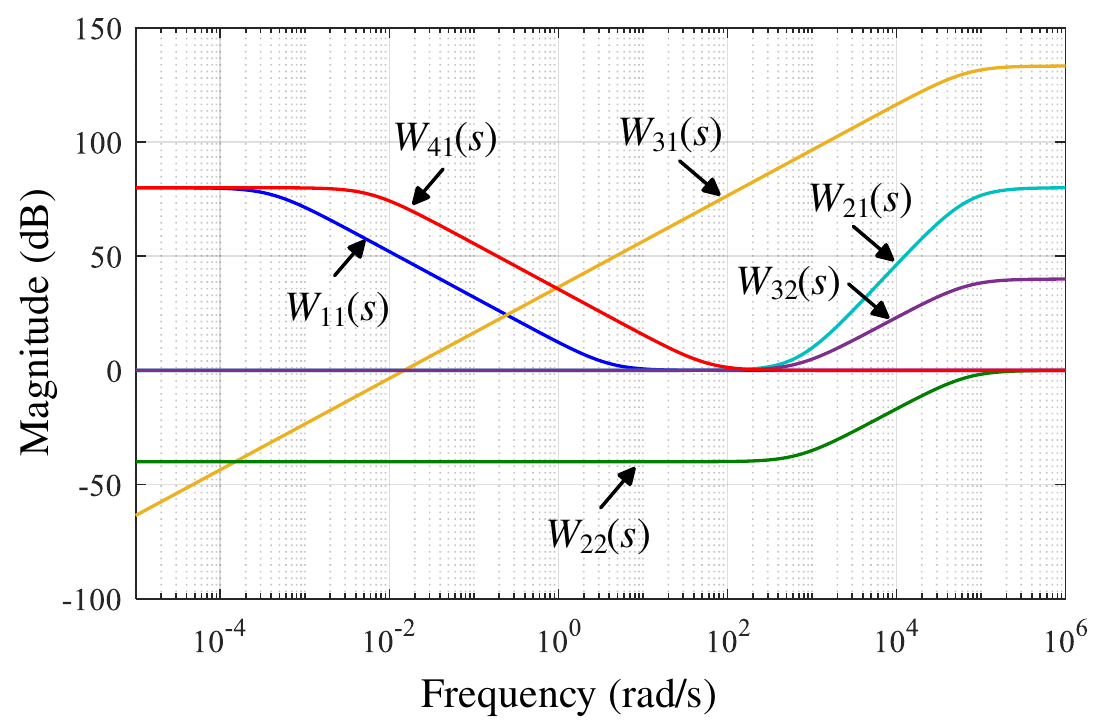}
	\caption{Log-magnitude curves of used weighting functions.}
	\label{weighting}
\end{figure}

The log-magnitude curves of the used $W_{ij}(s)$ are shown in Fig. \ref{weighting}. Afterwards, the gain vector $\bm K$ can be derived by solving (e.g., via Matlab's instructor \textit{hinfstruct} \cite{Gahinet2011})
\begin{align}
\label{h_infinity}
\min_{\bm K}||diag(W_{ij}(s)T_{ij}(s))||_\infty
\end{align}

\section{Simulation and Experimental Results}
\subsection{Simulation results}
To present the advantages of the proposed MIMO-GFM converter, the time-domain simulations results carried out by Matlab/Simulink are provided, where the topology is the same as in Fig. \ref{VSC}, and the parameters are listed in Table \ref{parameter}. The $\pazocal{H}_{\infty}$ optimization of (\ref{h_infinity}) is solved by the instructor \textit{hinfstruct} of Matlab with the initial values $\bm K=diag(90,400,0,0,0,20,0,0,0,0,0,1,1)$, which yields
\begin{align}
\bm\Phi=\left[
	\begin{matrix}
	120.224+\frac{265.6217}{s}&-0.0019&0&0.1673&-0.8274\\
	-0.8382&\frac{0.017622}{s+1.7622}&0&0&0\\
	-4.8977&0&0&\frac{1.0844}{s}&\frac{21.6872}{s}
	\end{matrix}
\right].
\end{align}

\begin{table}[!t]
	\renewcommand{\arraystretch}{1.3}
	\caption{Parameters Used in Simulations}
	\centering
	\label{parameter}
	\resizebox{\columnwidth}{!}{
		\begin{tabular}{c l c}
			\hline\hline \\[-3mm]
			Symbol & Description & Value  \\ \hline
			$\omega_n$  & nominal frequency & $100\pi$ rad/s \\
			$S_n $ & nominal power &  4 kW  \\ 
			$V_n $ & nominal line-to-line RMS voltage & 380 V \\
			$\omega_g $ & grid frequency & 1 p.u.  \\
			$V_g $ & grid voltage & 1 p.u. \\
			$L_g$  & line inductor & 2 mH \\
			$R_g$  & line resistor & 0.06 $\Omega$ \\
			$C_f$ & filter capacitor & 20 $\mu$F \\   
			$L_f$ & filter inductor & 2 mH \\
			$R_f$  & filter resistor & 0.06 $\Omega$ \\
			$C_{dc}$ & DC capacitor & 500 $\mu$F \\ 
			$D_p$ & droop coefficient of $P$-$f$ regulation & 0.01 p.u. \\ 
			$D_q$ &droop coefficient of $Q$-$V$ regulation & 0.05 p.u.\\
			$P_{ref}$ & Active power reference & 0.5 p.u.\\
			$Q_{ref}$ & Reactive power reference & 0 p.u.\\
			$V_{ref}$ & Voltage magnitude reference & 1 p.u.\\
			$V_{dcref}$ &DC voltage reference & 700 V\\[1.4ex]
			\hline\hline
		\end{tabular}
	}
\end{table}

To show the benefits of the added coupling terms, the VSG-2 and droop-5 controls in Table \ref{table_matrix} are used for comparison. The sparse control transfer matrix of VSG-2 is also designed by the same optimization routine as in (\ref{h_infinity}) as
\begin{align}
\bm\Phi=\left[
	\begin{matrix}
	90+\frac{400}{s}&0&0&0&0\\
	0&\frac{0.059801}{s+5.9801}&0&0&0\\
	0&0&0&\frac{1.9048}{s}&\frac{38.0954}{s}
	\end{matrix}
\right].
\end{align}
Meanwhile, the parameters of droop-5 are designed by its equivalent relationship with VSG-2 as shown in \cite{Liu2016a} and \cite{DArco2014}.

\begin{figure}[!t]
	\centering
	\includegraphics[width=0.8\columnwidth]{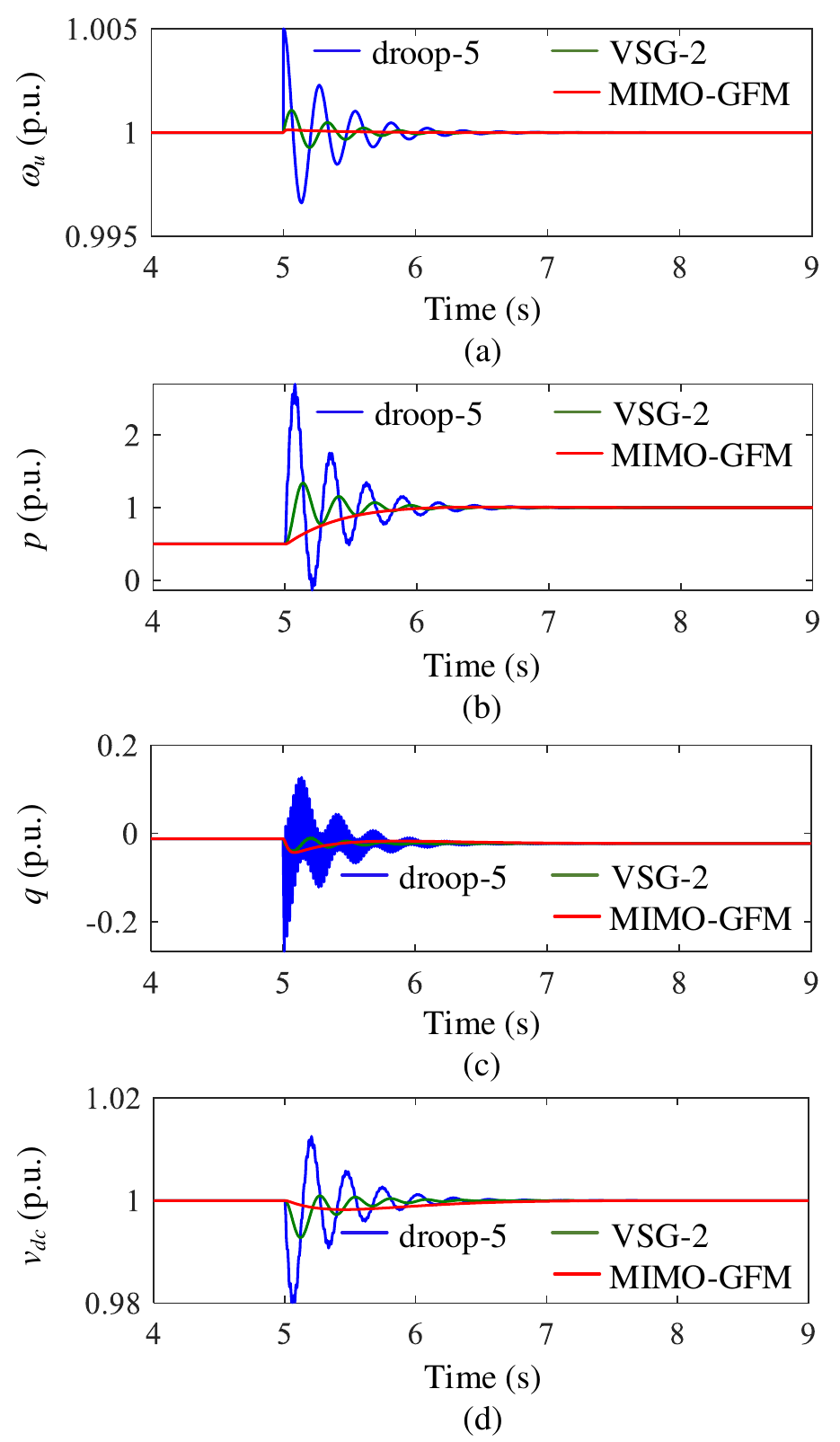}
	\caption{Simulation comparisons among different grid-forming converters when $P_{ref}$ steps from 0.5 p.u. to 1 p.u.}
	\label{simulation_pref}
\end{figure}

Fig. \ref{simulation_pref} presents the simulation comparisons when $P_{ref}$ steps from 0.5 p.u. to 1 p.u. As it is seen, both the droop-5 control and the VSG-2 control show larger oscillations responding to the disturbance. On the contrary, the proposed MIMO-GFM control can well damp this oscillation and has much smoother dynamics. It is also observed that the droop-5 control and the VSG-2 control have different response because $P_{ref}$ is seen as a disturbance at the input side as mentioned before.

\begin{figure}[!t]
	\centering
	\includegraphics[width=0.8\columnwidth]{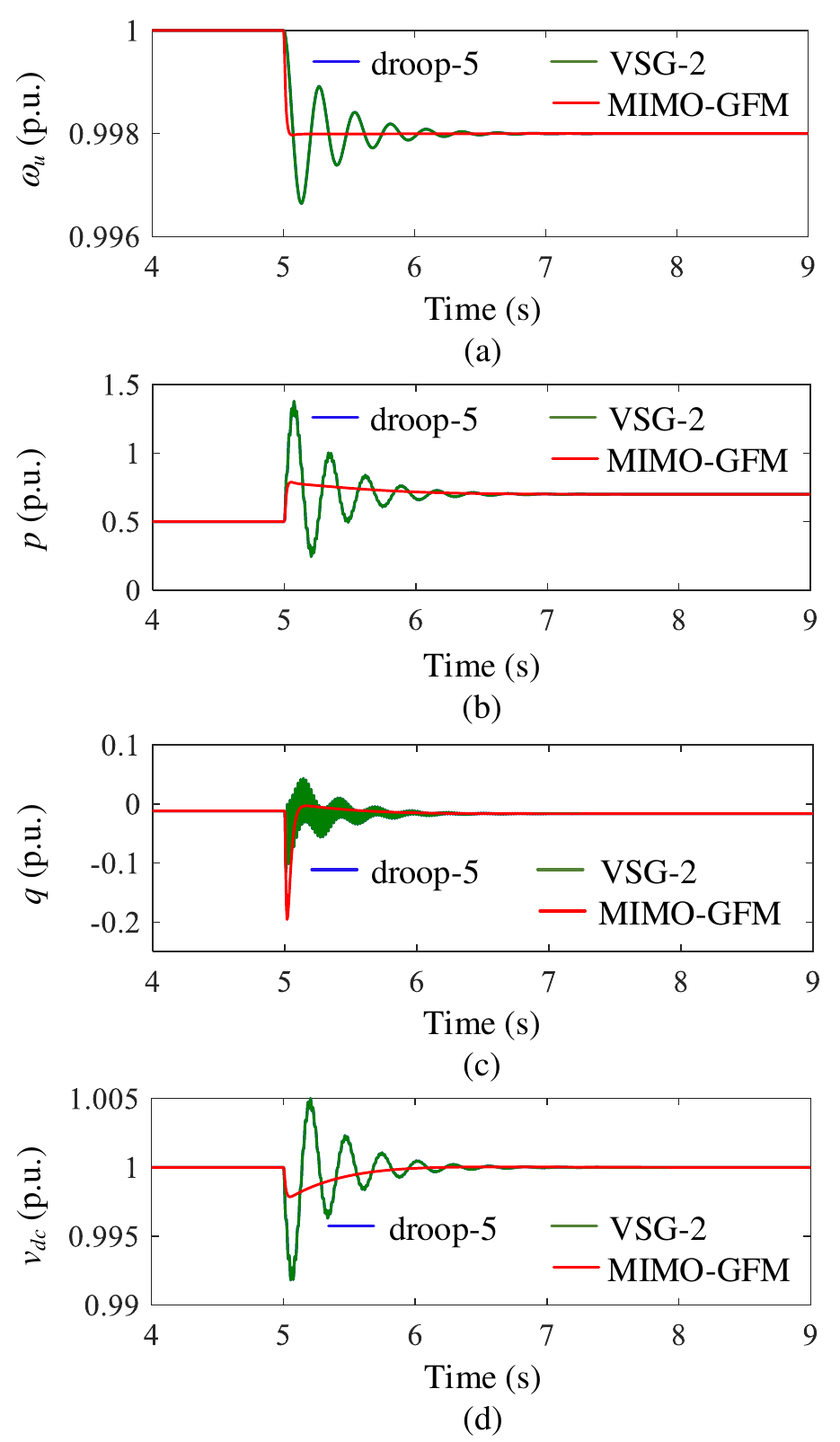}
	\caption{Simulation comparisons among different grid-forming converters when grid frequency decreases from 50 Hz to 49.9 Hz.}
	\label{simulation_wg}
\end{figure}

Fig. \ref{simulation_wg} presents simulation comparisons when $\omega_g$ decreases from 50 Hz to 49.9 Hz. As seen, the frequency of the proposed MIMO-GFM converter can synchronize to the power grid quickly and smoothly with almost no overshoot. In comparison, the droop-5 control and the VSG-2 control will lead to larger oscillations. It also shows the equivalence between the droop-5 control and the VSG-2 control in this case as $\omega_g$ is seen as a disturbance at the output side.

\subsection{Experimental results}
\begin{figure}[!t]
	\centering
	\includegraphics[width=\columnwidth]{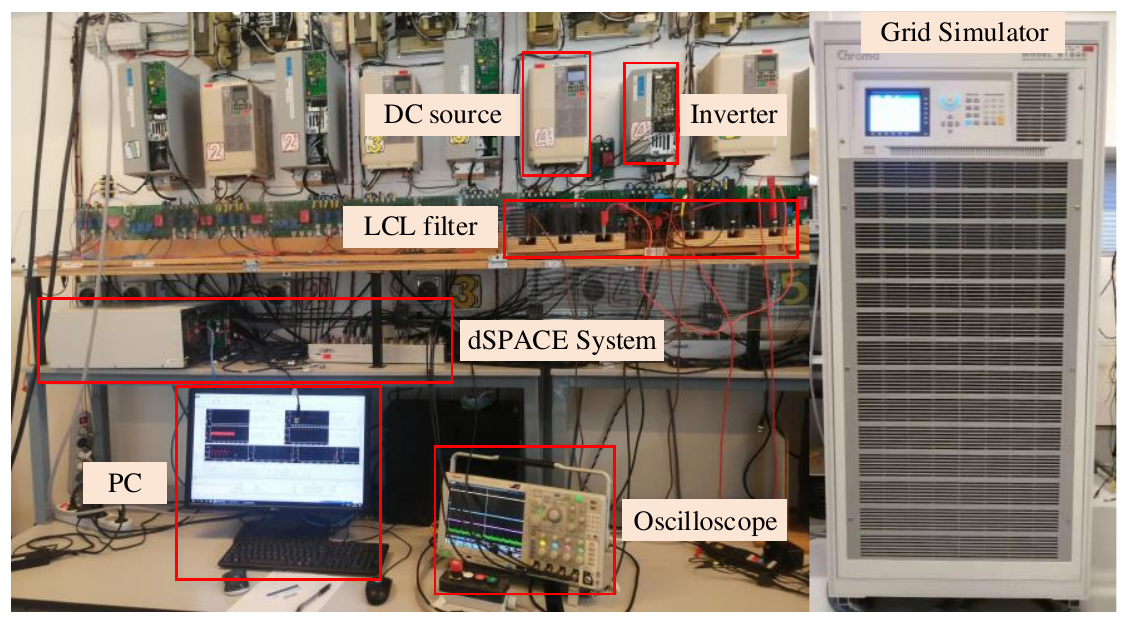}
	\caption{Experimental setup.}
	\label{setup}
\end{figure}

To further validate the effectiveness of the proposed MIMO-GFM control, an experimental setup is established, as shown in Fig. \ref{setup}. The power stage consists of a Danfoss drives system, an LCL filter and a Chroma 61845 grid simulator, while the control is implemented by the DS1007 dSPACE system. Meanwhile, the DS2004 A/D board and DS2101 D/A board are used to do measurement and generate the output, respectively. It should be mentioned that, because the DC control is fixed in the DC source provided by a Yaskawa D1000 regenerative converter, the first row of the control transfer matrix corresponding to the DC voltage control is not a design variable in the experiments. Nevertheless, we can still show the functions of the coupling terms in the frequency and voltage control in the second and third rows. The parameters of the setup are identical with Table \ref{parameter}.

\begin{figure}[!t]
	\centering
	\includegraphics[width=\columnwidth]{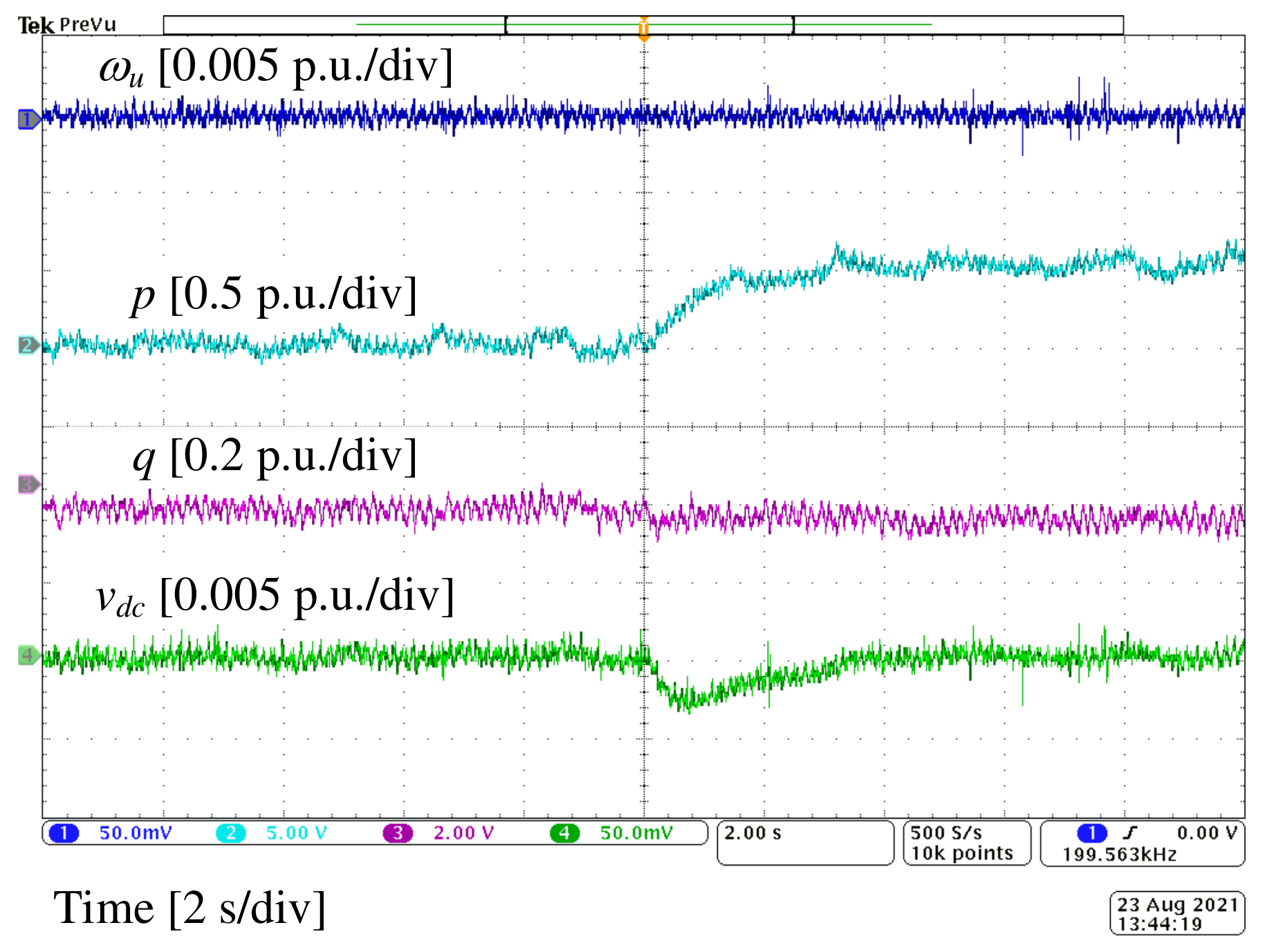}
	\caption{Experimental results of the proposed MIMO-GFM controller when $P_{ref}$ steps from 0.5 p.u. to 1 p.u.}
	\label{experiment_pref}
\end{figure}

\begin{figure}[!t]
	\centering
	\includegraphics[width=\columnwidth]{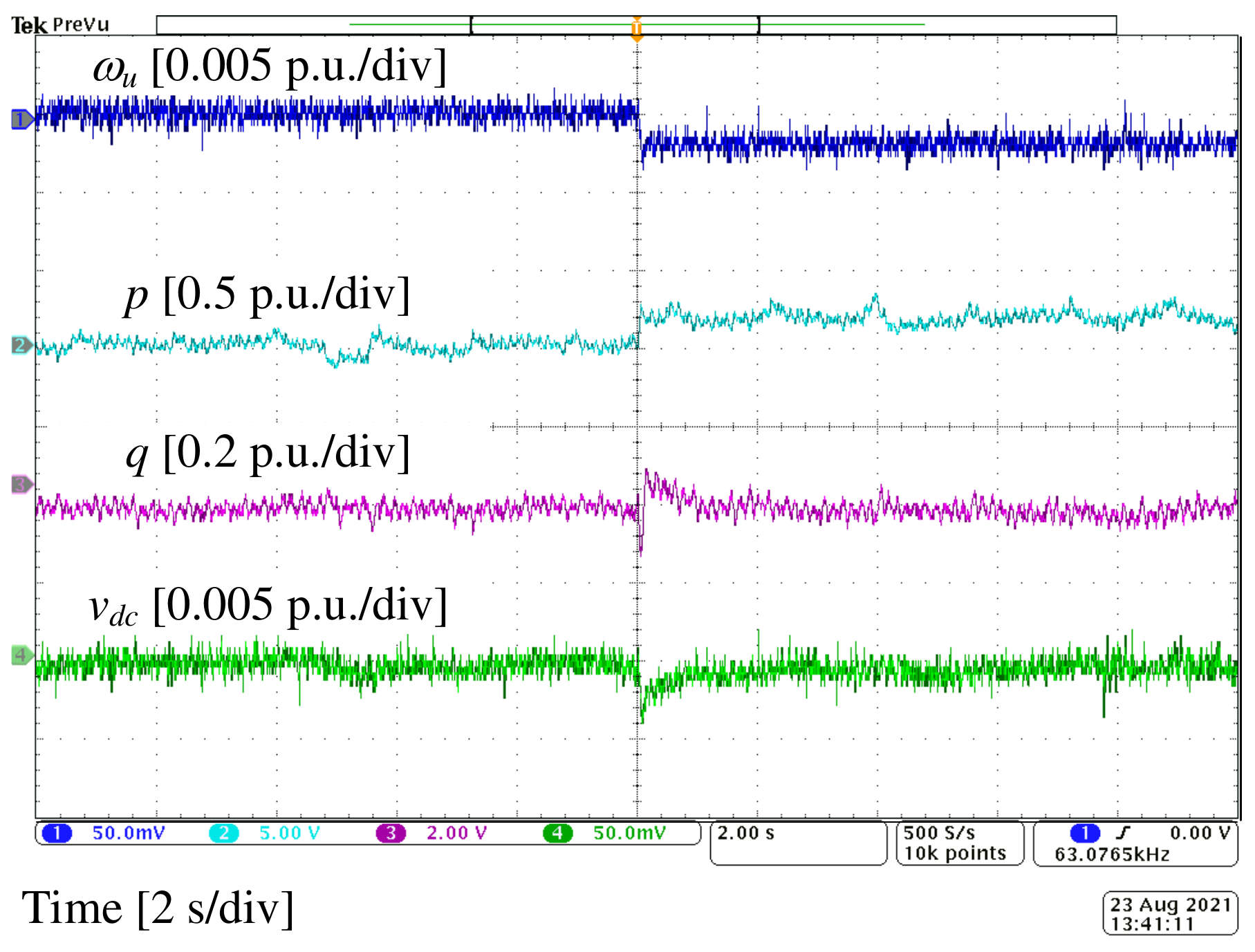}
	\caption{Experimental results of the proposed MIMO-GFM controller when grid frequency decreases from 50 Hz to 49.9 Hz.}
	\label{experiment_wg}
\end{figure}

\begin{figure}[!t]
	\centering
	\includegraphics[width=\columnwidth]{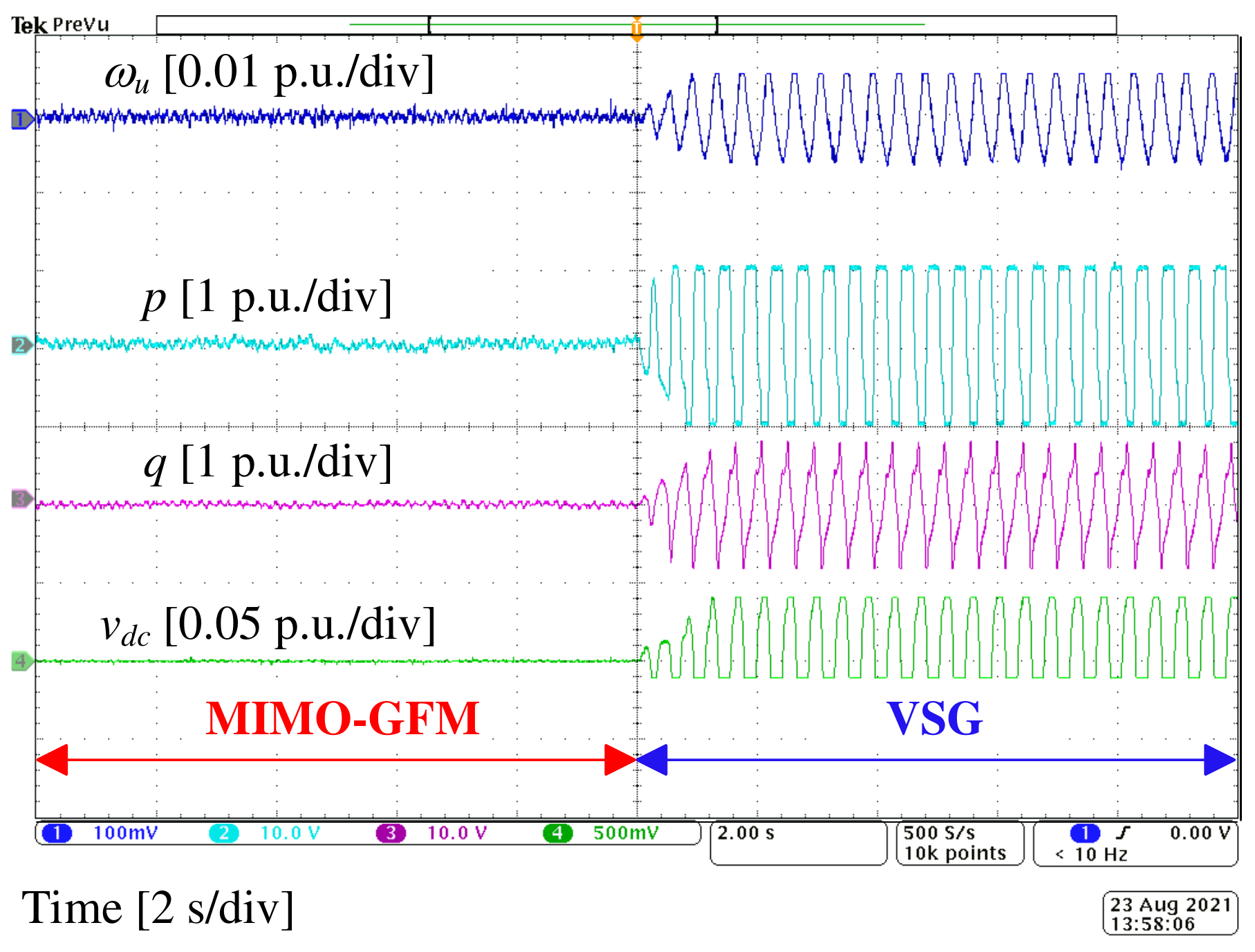}
	\caption{Experimental results when control strategy is switched from MIMO-GFM to VSG-2.}
	\label{experiment_VSG}
\end{figure}

Fig. \ref{experiment_pref} and Fig. \ref{experiment_wg} show the experimental results with the proposed MIMO-GFM control responding to the changes of $P_{ref}$ and $\omega_g$, respectively. As shown, the dynamics are fast and smooth with almost no oscillations, which validates a good performance of the proposed method. It should be mentioned that the traditional VSG control cannot be stabilized in the experiments, as shown in Fig. \ref{experiment_VSG}, due to various disturbances such as parameter errors and delays in the physically experimental system, although its oscillations can be damped in the simulations. This further highlights the robustness of the proposed method.

\section{Conclusion}
This paper proposes a generalized configuration for the grid-forming converter based on multi-input-multi-output feedback control theory. Instead of assuming that different loops are decoupled, the proposed configuration considers DC voltage control, frequency control, and voltage control as a single MIMO control transfer matrix to be designed. It is shown that many of the popular grid-forming controls as well as their improved formulations can be unified into a generalized control transfer matrix in the proposed configuration. Besides, this configuration is also helpful in comparison and design of controls. We also proposed a new MIMO-GFM control without increasing the order of the controller. To cope with the multiple control parameters, this paper presents how the optimal control design can be transformed to a standard $\pazocal{H}_{\infty}$ optimization problem. The simulation and experimental results verify the superior performance of the proposed method.


%



\ifCLASSOPTIONcaptionsoff
  \newpage
\fi



\bibliographystyle{IEEEtran}
\bibliography{IEEEabrv,TSG}
%

%








\end{document}